\documentclass[prl,aps,twocolumn,superscriptaddress,longbibliography,notitlepage,nobibnotes]{revtex4-1}

\usepackage[colorlinks,linkcolor=darkblue,citecolor=darkblue,urlcolor=darkblue]{hyperref}
\usepackage{microtype}
\usepackage{mathptmx}
\usepackage{amsmath}
\usepackage{amsfonts}
\usepackage{color}
\usepackage{graphicx}
\usepackage{float}
\definecolor{darkblue}{rgb}{0,0,0.6}
\newcommand{\To}{T_\textrm{o}}
\newcommand{\Tm}{T_\textrm{m}}
\newcommand{\Tf}{T_\textrm{f}}
\newcommand{\beq}{\begin{equation}}
\newcommand{\eeq}{\end{equation}}

\usepackage{soul}
\usepackage[normalem]{ulem}

\definecolor{deepblue}{rgb}{0,0,0.5}
\definecolor{deepred}{rgb}{0.6,0,0}
\definecolor{snapyellow}{RGB}{204,188,41}
\definecolor{snapred}{RGB}{204,41,41}
\definecolor{apsblue}{rgb}{0.18,0.19,0.57}
\hypersetup{
colorlinks,%
citecolor=deepblue,%
filecolor=black,%
linkcolor=deepblue,%
urlcolor=deepblue
}

\newcommand{\cancel}[1]{}

\graphicspath{{../figures/}:{./}}

\begin{document}

\title{Freezing, melting and the onset of glassiness in binary mixtures}

\author{Daniele Coslovich}
\email[Corresponding author: ]{dcoslovich@units.it}
\affiliation{Dipartimento di Fisica, Universit\`a di Trieste, Strada Costiera 11, 34151, Trieste, Italy}
\thanks{This article may be downloaded for personal use only. Any other use requires prior permission of the author and AIP Publishing. This article appeared in The Journal of Chemical Physics and may be found at \url{https://doi.org/10.1063/5.0252877}.}

\author{Leonardo Galliano}
\affiliation{Dipartimento di Fisica, Universit\`a di Trieste, Strada Costiera 11, 34151, Trieste, Italy}

\author{Lorenzo Costigliola}
\affiliation{``Glass and Time'', IMFUFA, Department of Science and Environment, Roskilde University, P.O. Box 260, DK-4000 Roskilde, Denmark}

\date{\today}

\begin{abstract}
  We clarify the relationship between freezing, melting, and the onset of glassy dynamics in a prototypical glass-forming mixture model.
  Our starting point is a precise operational definition of the onset of glassiness, as expressed by the emergence of inflections in time-dependent correlation functions.
  By scanning the temperature-composition phase diagram of the mixture, we find a disconnect between the onset of glassiness and freezing.
  Surprisingly, however, the onset temperature closely tracks the melting line, along which the excess entropy is approximately constant.
  At fixed composition, all characteristic temperatures display nonetheless similar pressure dependencies, which are very well predicted by the isomorph theory.
  While our results rule out a general connection between thermodynamic metastability and glassiness, they call for a reassessment of the role of crystalline precursors in glass-forming liquids.
\end{abstract}

\maketitle
\date{\today}

Two-step relaxation is one of the most prominent features of {liquids} approaching
the glass transition~\cite{cavagnaSupercooledLiquidsPedestrians2009}.
It manifests itself as a plateau in time-dependent correlation functions and reflects the strong separation of timescales between fast microscopic motion and a spectrum of much slower relaxation processes.
In a thermal cooling process, two-step relaxation first appears around a crossover temperature, $\To$, that marks the onset of slow dynamics~\cite{sastrySignaturesDistinctDynamical1998} or, as we shall write in the following, ``onset of glassiness''~\cite{cavagnaSupercooledLiquidsPedestrians2009}.
Below $\To$, the system ceases to be a normal liquid and develops the typical features of glassy dynamics, including dynamic heterogeneity and super-Arrhenius dependence of relaxation times.
Predicting and explaining the origin of these features in materials as diverse as supercooled liquids~\cite{berthierTheoreticalPerspectiveGlass2011}, spin glasses~\cite{youngSpinGlassesRandom1998} or type-II superconductors~\cite{nattermannVortexglassPhasesTypeII2000} is a key challenge in the field of disordered systems.

One important and yet often eluded question is whether the onset of glassiness is related or not to thermodynamic metastability with respect to an ordered phase.
From a theoretical standpoint, a connection between the onset of glassiness and frustrated crystallization has been invoked in a phenomenological model of the glass transition~\cite{tanakaTwoorderparameterDescriptionLiquids1999}.
In a similar vein, the so-called frustration-limited domains theory~\cite{tarjusfrustrationbasedapproachsupercooled2005} attributes the onset of glassiness to an avoided crystallization, which occurs however in a curved space where geometric frustration is lifted.
By contrast, in theoretical approaches based on mean-field spin glass models or infinite-dimensional systems~\cite{parisiTheorySimpleGlasses2020} crystallization is ruled out from the outset, and the onset of glassiness results from activated transitions between amorphous metastable states~\cite{brumerMeanfieldTheoryModecoupling2004, biroliRandomFirstOrderTransition2012}.
Most other theories of the glass transition~\cite{berthierTheoreticalPerspectiveGlass2011} also neglect the role of crystallization.

From an empirical point of view, it is clear that thermodynamic metastability is not a necessary condition for glassy dynamics: liquid silica, for instance, displays glassy dynamics already at the melting temperature $\Tm$~\cite{horbachStaticDynamicProperties1999}.
A connection between crystallization and the onset of glassiness may nonetheless hold within a subset of liquids~\cite{greetEmpiricalCorrespondingstatesRelationship1967}.
Previous analysis of experimental data~\cite{tanakaTwoorderparameterDescriptionLiquids1999a} indicated a rough correlation between the melting temperature $\Tm$ and a crossover temperature $T_A$, akin to $\To$, at which super-Arrhenius behavior sets in~\cite{hansenDynamicsGlassformingLiquids1997}.
Recent simulation studies of a Lennard-Jones (LJ) binary mixture even suggested an identity, $\Tm\approx \To$~\cite{banerjeeDeterminationOnsetTemperature2017, pedersenPhaseDiagramKobAndersenType2018}.
In multi-component systems, however, the liquid phase can be metastable with respect to a pure crystal, to coexisting liquid and crystal phases or to phase-separated crystals, and freezing and melting transitions must be clearly distinguished.
These aspects have not been addressed in any detail so far.
There are also indications that thermodynamic properties related to the excess entropy~\cite{bellExcessentropyScalingSupercooled2020} change sharply around $\To$~\cite{banerjeeDeterminationOnsetTemperature2017}.
Our goal is to clarify these connections in a prototypical glassy binary mixture with different chemical compositions, disentangling the role of freezing and melting, and to provide the basis for a systematic analysis on a broader range of liquids.

One technical hurdle is the determination of the onset temperature $\To$, for which there is no generally agreed, operational definition.
In fact, while the concept of the onset of glassiness is well-established, its practical determination has never been standardized.
The appearance of the super-Arrhenius dependence of the structural relaxation time $\tau_\alpha$ can be used as a proxy for $\To$, but such a procedure is often done ``by eye'' or assuming a functional form for $\tau_\alpha(T)$~\cite{hansenDynamicsGlassformingLiquids1997, tarjusfrustrationbasedapproachsupercooled2005, coslovichUnderstandingFragilitySupercooled2007b}. 
The onset of glassiness is also associated with a decrease in the inherent structure energy with decreasing temperature~\cite{sastrySignaturesDistinctDynamical1998, brumerMeanfieldTheoryModecoupling2004}, but the crossover is broad and does not provide a clear-cut definition.
Alternative but more complex procedures have been proposed~\cite{schoenholzStructuralApproachRelaxation2016, singhEmergenceCooperativelyReorganizing2021}.
In the following, we provide a straightforward and precise definition of the onset of glassiness in terms of the inflection points of time-dependent correlation functions.
Our procedure removes the above difficulties and is generally applicable to both simulation and experimental data, as it does not require knowledge of the interaction potential.

In this work, we study the composition dependence of the onset of glassiness in the Kob-Andersen (KA) mixture~\cite{kobTestingModecouplingTheory1995a}, which is an LJ mixture of two types of particles, $A$ and $B$.
The concentration of $B$ particles is $x$, with $x=0.2$ corresponding to the canonical KA mixture.
We performed molecular dynamics simulations for a KA mixture composed of $N=500$ particles using the \texttt{atooms} simulation framework~\cite{atooms}, both along constant pressure and constant density paths.
As a cross-check, we did additional simulations for a larger system size ($N=8000$) using \texttt{RUMD}~\cite{baileyRUMDGeneralPurpose2017}.
All quantities are expressed in standard LJ units.
More details are given in the Supplementary Material (SM)~\cite{si}.

We compute the self intermediate scattering function $F_s(k,t)$ for the $A$-particles (the results for the $B$-particles are qualitatively similar~\cite{si}).
We focus on wavenumbers $k$ around the first peak of the structure factor $S_{AA}(k)$.
The calculation of the time derivative $F_s^\prime(k,t)$ requires some care as it can be affected by statistical noise: 
to cope with this, we first compute $dF_s(k,t)/d\log{t} = tF_s^\prime(k,t)$ by central differences, then fit it to a fourth order polynomial in $\log{t}$, restricting the time range such that $F_s$ is between 0.05 and 0.98, and extract its critical points.
Small changes to the fit range do not change our conclusions.
The onset is defined by the emergence of two minima in $tF_s^\prime(k,t)$.
The height $\delta$ of the smallest barrier separating the two minima of $tF_s^\prime$ provides a precise, observable-dependent order parameter for the onset of glassiness: $\delta$ becomes finite below $\To$.
Although $\To$ depends, in principle, on the chosen observable, we found that its dependence on wavenumber, chemical species and correlation function is weak; see below.
Therefore, as a crossover, the onset of glassiness remains well-defined.

\begin{figure}
  \includegraphics[width=\linewidth]{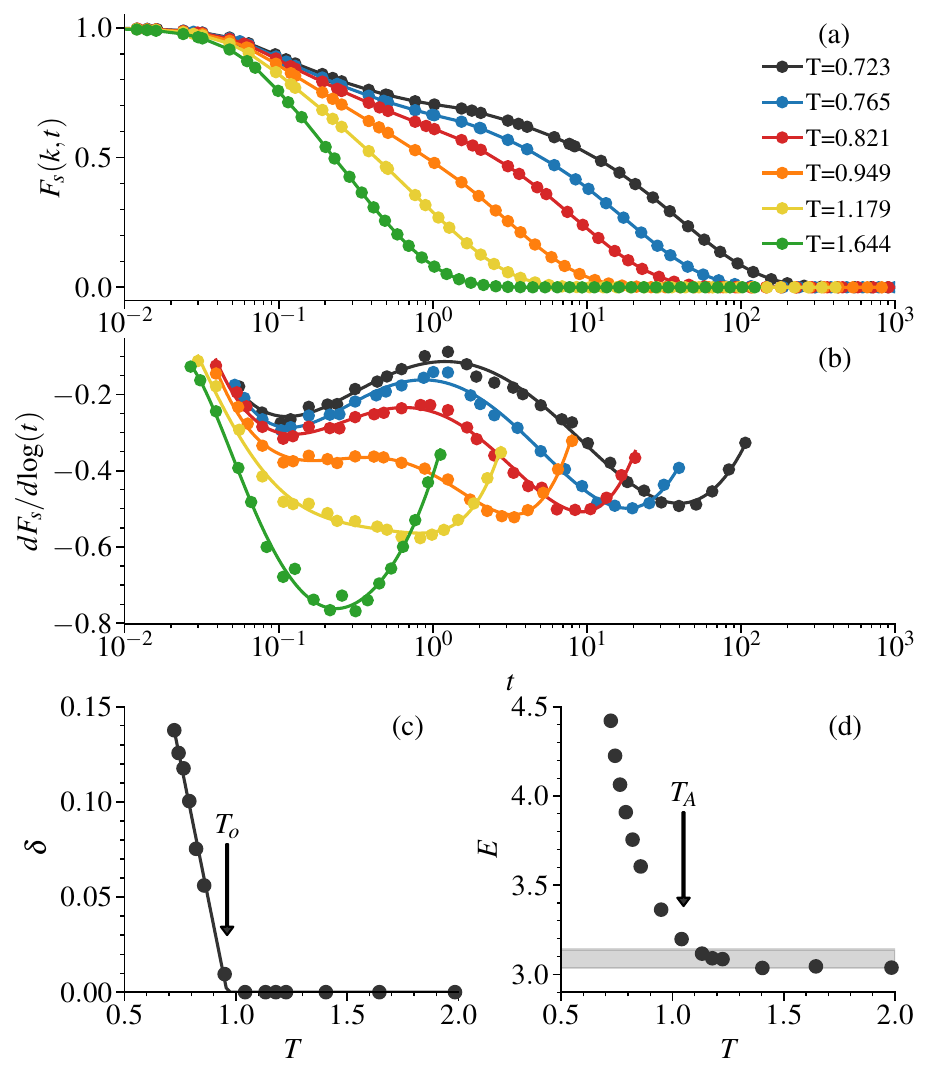}
  \caption{Protocol to define the onset temperature, illustrated along the isobar $P=10.19$ for $x=0.2$:
    (a) $F_s(k,t)$ of $A$-particles ($k=7.2$) for temperatures around $\To$;
    (b) $tF_s^\prime(k,t)$ for the same temperatures as in (a), the solid lines are fourth order polynomial fits;
    (c) order parameter $\delta(T)$ obtained from $tF_s^\prime(k,t)$, the solid line is a linear fit vanishing at $\To$;
    (d) activation energy $E(T)$.
  In (c) and (d), the arrows mark $\To$ and $T_\mathrm{A}$, respectively. In (d), the shaded area indicates the estimated range of validity of the high-temperature Arrhenius fit.
}
  \label{fig:overview}
\end{figure}

Our procedure is illustrated in Fig.~\ref{fig:overview} for the canonical concentration $x=0.2$ at pressure $P=10.19$ and wavenumber $k=7.2$.
We start by equilibrating the system at a high temperature, in the liquid phase, and decrease the temperature at fixed pressure until $\delta>0$.
Once $\To$ has been bracketed, we perform additional simulations to refine its determination.
Panel (a) displays $F_s(k,t)$ around $\To$.
The definition of the onset can be grasped from panels (b) and (c): below $\To$, the time derivative $tF_s^\prime(k,t)$ displays one local maximum, corresponding to the sought-for inflection of $F_s(k,t)$.
The polynomial fit, indicated by solid lines, is fairly robust and works well even with noisy data~\cite{si}.
Small oscillations in $tF_s^\prime(k,t)$ at short times, unrelated to the onset of glassiness~\cite{Brazhkin_Fomin_Lyapin_Ryzhov_Tsiok_Trachenko_2013}, are visible with high-quality statistics, but they do not affect the calculation of $\delta$~\cite{si}.
To locate $\To$ precisely, we fit $\delta(T)$ to a linear function that vanishes at $\To$.
A simpler bisection procedure, which avoids any fitting, provides consistent results~\cite{si}.
Finally, we show in panel (d) the activation energy $E(T)=T \log(\tau_\alpha/\tau_\infty)$~\cite{tarjusfrustrationbasedapproachsupercooled2005} associated with the structural relaxation time $\tau_\alpha(T)$, at which $F_s(k,\tau_\alpha)=1/e$.
The microscopic time $\tau_\infty$ is obtained by fitting the high-temperature data to an Arrhenius equation.
As expected, the onset of two-step relaxation is accompanied by the appearance of super-Arrhenius temperature dependence at $T_A$.
We find that $T_A$ is slightly higher than $\To$ 
and is sensitive to the range of the Arrhenius fit, which holds anyway only approximately at high temperature~\cite{costigliolaCommunicationSimpleLiquids2018}.

\begin{figure}
   \includegraphics[width=\linewidth]{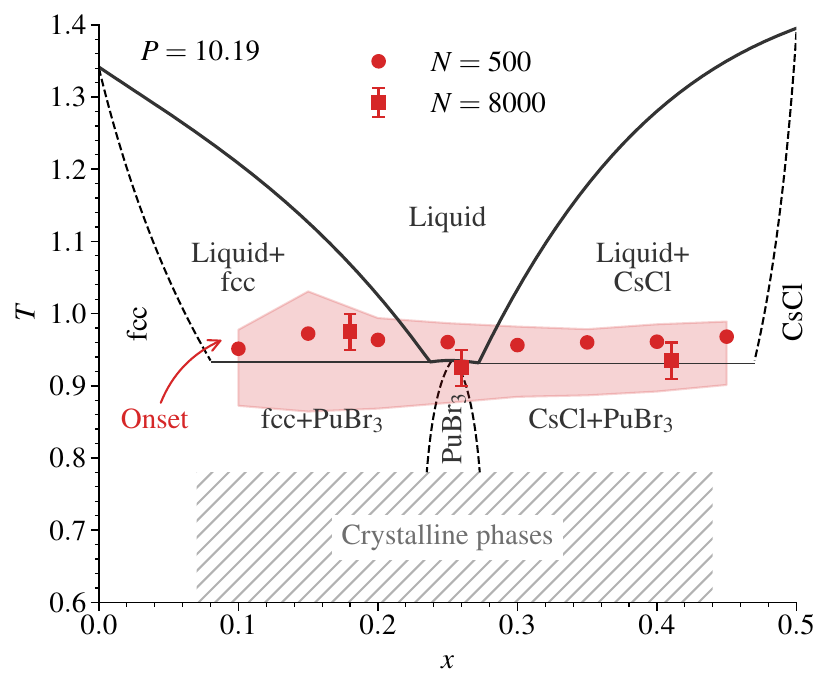}
   \caption{Onset temperatures $\To$ for $N=500$ (circles) and $N=8000$ (squares) as a function of $x$ at $P=10.19$.
     The shaded area indicates the range of $\To$ obtained for values of $k$ in the interval between $5$ and $10$.
     The lines are the phase boundaries inferred from Ref.~\onlinecite{pedersenPhaseDiagramKobAndersenType2018}.
     The thick and thin solid lines indicate the freezing and melting lines, respectively.
     The dashed lines represent estimates of the stability limits of the {solid solutions}.
    }
  \label{fig:phase_diagram}
\end{figure}

We now turn to the relation between the onset of glassiness and crystallization.
In Ref.~\onlinecite{pedersenPhaseDiagramKobAndersenType2018}, Pedersen \textit{et al.} studied the phase diagram of the KA mixture and measured the freezing (or ``liquidus'') line, below which the liquid crystallizes at least partially~\footnote{In Ref.~\onlinecite{pedersenPhaseDiagramKobAndersenType2018} the freezing line was referred to as ``melting line''.}.
The stable crystalline phases for $x\rightarrow 0 $ and $x\rightarrow 0.5$ are fcc and CsCl, respectively.
At the fcc-CsCl eutectic composition $x=0.25$, however, the stable crystalline phase has a PuBr$_3$ symmetry~\cite{pedersenPhaseDiagramKobAndersenType2018} and coexists with the liquid in a narrow range of compositions.
{As shown in the SM~\cite{si}, this implies the presence of two very close eutectic points.}
We reproduce these results for $P=10.19$ as thick solid lines in Fig.~\ref{fig:phase_diagram}.
{From the location of the eutectic points and} using the Gibbs phase rule~\cite{silbeyPhysicalChemistry2004, si}, we infer 
the melting (or ``solidus'') lines of the underlying {phase-separated} crystals: above these lines, the ordered phases melt at least partially.
The phase diagram at low temperatures is instead unknown.
We superpose on the phase diagram the onset temperatures obtained from the protocol described above.
The shaded area indicates the range of onsets corresponding to $k$ values around the first peak of $S_{AA}(k)$, i.e., from $k=5$ to $10$~\cite{si}, while the filled points are for $k=7.2$.
It is clear that $\To$ does not track the freezing temperature $\Tf$: while the latter displays the typical V-shape of eutectic mixtures, $\To$ does not show any systematic variation with $x$.
Similar results are observed for the $B$-particles and when using correlation functions probing collective dynamics~\cite{si}.
The corresponding onset temperatures lie within the shaded area indicated in Fig.~\ref{fig:phase_diagram}.

The qualitatively different trends of $\To$ and $\Tf$ demonstrate that the onset of glassiness and freezing are disconnected: 
the liquid can be thermodynamically metastable without displaying glassy dynamics.
Surprisingly, however, the onset of glassiness closely follows the melting line, which runs horizontally in the phase diagram. 
At this stage, we emphasize that the correspondence of $\To$ with the melting of the underlying stable crystals could be coincidental.
Proving a causal connection would require a direct determination of the crystalline precursors in the metastable liquid -- we will come back to this point in the closing paragraphs.

The results shown in Fig.~\ref{fig:phase_diagram} also allow us to dissipate a possible source of confusion.
Analyzing data at $x=0.2$, Pederesen \textit{et al.}~\cite{pedersenPhaseDiagramKobAndersenType2018} found an ``identity between onset and melting temperatures'', with the former defined as the appearance of super-Arrhenius behavior~\cite{coslovichUnderstandingFragilitySupercooled2007b}.
We note, however, that the word melting was used in Ref.~\onlinecite{pedersenPhaseDiagramKobAndersenType2018} to indicate freezing.
As we can see from Fig.~\ref{fig:phase_diagram}, $\Tm \approx \Tf \approx \To$ around the eutectic compositions, but deviations are found for other values of $x$.
An approximate identity between $\To$ and $\Tm$ holds in the KA mixture, but for a different reason from the one implied by Ref.~\onlinecite{pedersenPhaseDiagramKobAndersenType2018}.
The absence of a direct connection between $\Tf$ and $\To$ may be partly inferred from the trends of the iso-diffusivity lines shown in Ref.~\onlinecite{pedersenPhaseDiagramKobAndersenType2018}, which only show a weak, monotonic dependence on $x$.
We obtained indeed similar results for the relaxation time itself~\cite{si}.
However, transport coefficients \textit{per se} do not provide direct information about the shape of time-dependent correlation functions.

\begin{figure}
  \includegraphics[width=\linewidth]{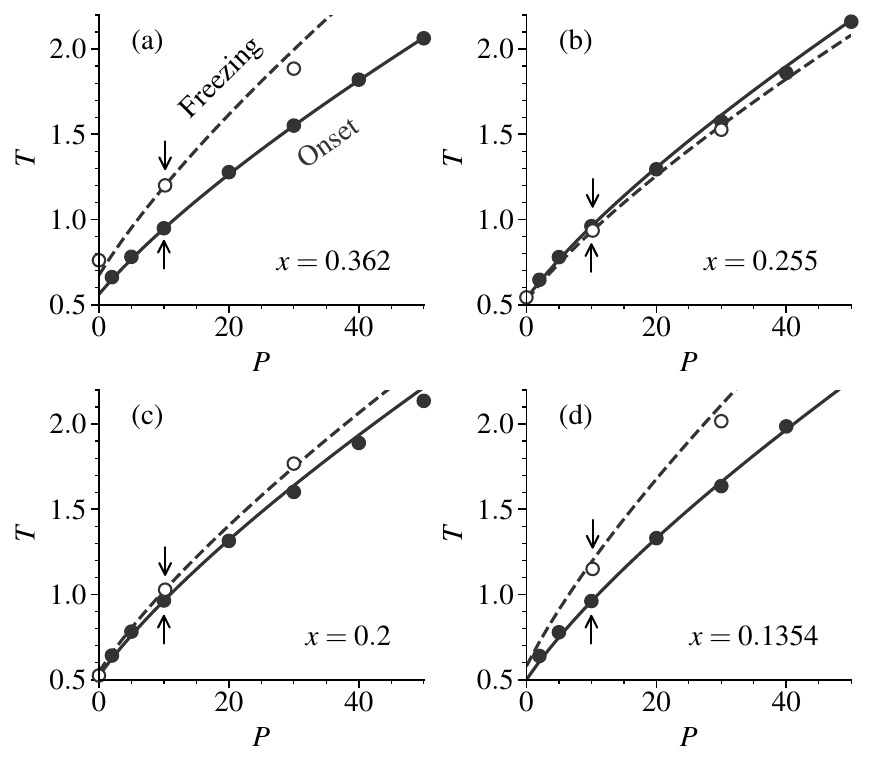}
  \caption{Pressure dependence of the freezing temperature $\Tf$ (open circles) and of the onset temperature $\To$ (filled circles) for (a) $x=0.362$, (b) $x=0.255$, (c) $x=0.2$ and (d) $x=0.1354$.
    The theoretical predictions of the isomorph theory for $\Tf$ and $\To$ are shown as dashed and solid lines, respectively.
    The arrows mark the reference states used for the predictions.
  }
  \label{fig:pressure}
\end{figure}

Even though the freezing line does not track the onset of glassiness in the $T$-$x$ diagram, $\Tf$ and $\To$ scale similarly as a function of pressure.
This is shown in Fig.~\ref{fig:pressure} for selected compositions.
We used a constant wavenumber $k=7.2$ for the calculation of the onset temperature, independent of composition and pressure.
Interestingly, it is possible to accurately predict the pressure dependence of both quantities using the isomorph theory~\cite{dyrePerspectiveExcessentropyScaling2018}.
{By assuming that they both follow an isomorph, we predict $\Tf(P)$ and $\To(P)$ at any fixed $x$ from the sole knowledge of thermodynamic properties at $\Tf(P_0=10.19)$ and $\To(P_0=10.0)$, respectively.
See Refs.~\cite{pedersenPhaseDiagramKobAndersenType2018, schroderPressureenergyCorrelationsLiquids2011, ingebrigtsenCommunicationThermodynamicsCondensed2012} and the SM for more details.}
The agreement is excellent for $\To$, while some discrepancies are seen for $\Tf$ at $x=0.362$, see also Ref.~\onlinecite{pedersenPhaseDiagramKobAndersenType2018}.
This is consistent with the observation that the freezing line is an isomorph only approximately~\cite{pedersenThermodynamicsFreezingMelting2016}.

{Our results indicate that the changes in the onset temperature are closely connected to those of the excess entropy $S_\textrm{ex}$, which is defined as the difference between the total entropy and its ideal gas contribution~\cite{singhStructuralCorrelationsCooperative2012a, bellExcessentropyScalingSupercooled2020} and is constant along an isomorph.
Previous studies have also connected $\To$ to a change in the $n$-body
contributions to the excess entropy per particle $s_\textrm{ex}=S_\textrm{ex}/N$.}
More precisely, the residual many-body entropy is defined as $\Delta s = s_\textrm{ex} - s_2$, where $s_2$ is the two-body approximation to $s_\textrm{ex}$~\cite{baranyaiDirectEntropyCalculation1989}.
It was found that $\Delta s$ changes sign, in the KA mixture, at a temperature slightly lower than $\To$~\cite{singhStructuralCorrelationsCooperative2012a}.
Note that $\Delta s = 0$ has also been proposed as an empirical criterion for freezing in one-component liquids~\cite{giaquintaEntropyFreezingSimple1992, krekelbergResidualMultiparticleEntropy2008a}.

To perform a stringent test of these ideas, we compute $s_\textrm{ex}$ and $s_2$ over a range of compositions.
We carry out thermodynamic integration from a low-density, high-temperature state ($\rho=10^{-4}$, $T=5$), where we assume that the system behaves like an ideal gas~\cite{singhStructuralCorrelationsCooperative2012a}.
To determine $s_\textrm{ex}$ along the isobar $P=10.19$, we first follow an isotherm at $T=5$ up to density $\rho(T=5, P=10.19)$ and then proceed with batches of small isothermal and isochoric paths, keeping the system along the selected isobar~\cite{si}.
We also compute $s_\textrm{ex}$ along isochores for each composition, fixing the density at $\rho(T=\Tf, P=10.19)$.
Our results for $s_\textrm{ex}$ agree within error bars with those of Bell \textit{et al.}~\cite{bellExcessentropyScalingSupercooled2020}.

\begin{figure}
  \includegraphics[width=\linewidth]{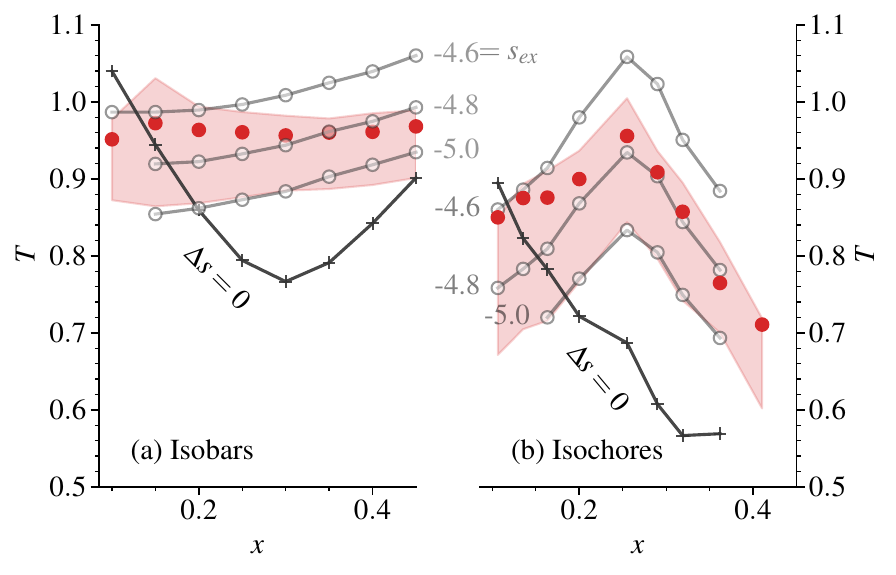}
  \caption{Excess entropy measures in the $T$-$x$ diagram along (a) isobars at $P=10.19$ and (b) isochores at densities corresponding to freezing, $\rho(T=\Tf, P=10.19, x)$.
    The temperatures at which $s_\textrm{ex}=\textrm{const}$ and $\Delta s=0$ are indicated as open empty circles and crosses, respectively.
  The onset temperatures are shown as in Fig.~\ref{fig:phase_diagram}.}
  \label{fig:entropy}
\end{figure}

Excess entropy data are shown in Fig.~\ref{fig:entropy}.
Interestingly, the lines of constant $s_\textrm{ex}$ display the same qualitative behavior as the onset of glassiness: they vary weakly with $x$ at constant pressure and are non-monotonic along isochoric paths, with a maximum around the eutectic composition{s}.
We also found that the reduced structural relaxation times for different $x$ collapse on a master curve when shown as a function of the excess entropy~\cite{si}, in line with the concept of quasi-universality put forth by Bell \textit{et al.}~\cite{bellExcessentropyScalingSupercooled2020}. 
The value of $s_\textrm{ex}$ at $\To$ lies between -4.7 and -4.9.
Combining these observations with those inferred from Fig.~\ref{fig:phase_diagram}, we conclude that melting occurs in this same range of excess entropies and could be a quasi-universal property in the sense of Ref.~\onlinecite{bellExcessentropyScalingSupercooled2020}.

From Fig.~\ref{fig:entropy} we also see that the locus of points where $\Delta s=0$ has a non-monotonic behavior, with a minimum around the eutectic compositions.
This trend is thus qualitatively similar to the one of the freezing line, cf. Fig.~\ref{fig:phase_diagram}, but $\Delta s$ vanishes at a temperature lower than $\Tf$ by 10--50\% depending on composition.
It is also clear that $\Delta s = 0$ does not provide a sound criterion for the onset of glassiness, which occurs, instead, around the same temperature irrespective of $x$.
Similar discrepancies are observed on paths at constant density, see panel (b):
the non-monotonic behavior of $\To$ is well reproduced by the lines of constant $s_\textrm{ex}$, while those at which $\Delta s=0$ display a qualitatively different trend.
Therefore, the splitting of $s_\textrm{ex}$ into two-body and many-body contributions does not bear any clear connection with the onset of glassiness in the KA mixture.

One conclusion to be drawn from our work is the lack of a general connection between thermodynamic metastability and the onset of glassiness.
Not only a liquid can be highly viscous without being metastable, as is the case for silica: the large gap between $\Tf$ and $\To$ seen in Fig.~\ref{fig:phase_diagram} shows that a liquid can be metastable without displaying yet any glassy feature.
Key to this observation is the distinction between freezing and melting, which was not duly taken into account in previous computational studies of the KA mixture.
The second main result is that the onset of glassiness occurs close to the solidus line, which marks the melting of the underlying crystalline phases.
The generality of this connection can now be tested straightforwardly for computational models whose phase diagram is known~\cite{yiannourakouPhaseEquilibriumColloidal2009a, russoGlassFormingAbility2018, molineroTuningTetrahedralitySilicon2006}.
We did preliminary calculations for the Wahnström LJ mixture, whose phase diagram has been determined recently~\cite{nishioLiquidusCurveLennardJones2024}, and we found very similar results to those presented therein: $\To$ and $\Tm$ are very close to each other.
Our method to determine $\To$ can, in principle, be adapted to the analysis of dynamic susceptibilities relevant to experiments~\cite{blochowiczSusceptibilityFunctionsSlow2003}, where precise measurements of the onset temperature are rare, see Ref.~\onlinecite{schmidtkeReorientationalDynamicsMolecular2013} for an exception.

The apparent correspondence between $T_m$ and $T_o$ motivates a critical reassessment of the role of locally favored structures in glass-forming liquids~\cite{royallRoleLocalStructure2015, tanakaRevealingKeyStructural2019, weiAssessingUtilityStructure2019}.
It has been argued that the competition between crystalline precursors corresponding to different crystalline phases can contribute to stabilize the metastable liquid~\cite{russoGlassFormingAbility2018}.
In this respect, it would be interesting to identify precursors of the PuBr$_3$ phase in the KA mixture, as this is the stable phase close to the eutectic compositions.
This crystalline phase contains bicapped trigonal prisms, akin to some of the locally favored structures of the model at the canonical composition~\cite{coslovichLocallyPreferredStructures2011}, and has been so far overlooked in the crystallization studies of the KA model~\cite{crowtherNatureGeometricFrustration2015, nandiCompositionDependenceGlass2016a, ingebrigtsenCrystallizationInstabilityGlassForming2019}.
We emphasize that the connection between melting and the onset of glassiness could be specific to liquids that ``borrow'' their local structure from the underlying stable crystal~\cite{pedersenHowSupercooledLiquid2021}.
Structural analyses across the melting line will provide a crucial test of the role of crystalline precursors.

{
  Predicting when liquids first start to show glassy behavior is a well-defined open problem, and currently a challenge for the theories of the glass transition that rely on structure or thermodynamics. 
Our work lays down the basis to address this problem quantitatively and short-lists some of the proposed solutions, starting from entropy-based approaches.
Among the outstanding approaches, we expect mode-coupling theory~\cite{naurothQuantitativeTestModecoupling1997, gotzeEffectCompositionChanges2003} and its extensions~\cite{luoGeneralizedModecouplingTheory2020a, ciarellaMulticomponentGeneralizedModecoupling2021} to provide at least qualitatively correct predictions.
Another promising approach to predict the onset of glassiness from first principles builds on the relationship between the {caging} dynamics and the local curvature of the potential energy surface, see Refs.~\onlinecite{sunGeneralStructuralOrder2022, coslovichRevisitingSinglesaddleModel2022} for recent work in this direction.
}





\section*{Data availability}
The data and workflow necessary to reproduce the findings of this study are available in the Zenodo data repository~\cite{zenodo}.

\section*{Acknowledgments} 
We thank Jeppe C. Dyre, Walter Kob and Ulf R. Pedersen for useful discussions.
LC acknowledges support from VILLUM Foundation \textit{``Matter''} grant (No. 16515).
The article has been produced with co-funding from the European Union - Next Generation EU.


\begin{thebibliography}{57}%
\makeatletter
\providecommand \@ifxundefined [1]{%
 \@ifx{#1\undefined}
}%
\providecommand \@ifnum [1]{%
 \ifnum #1\expandafter \@firstoftwo
 \else \expandafter \@secondoftwo
 \fi
}%
\providecommand \@ifx [1]{%
 \ifx #1\expandafter \@firstoftwo
 \else \expandafter \@secondoftwo
 \fi
}%
\providecommand \natexlab [1]{#1}%
\providecommand \enquote  [1]{``#1''}%
\providecommand \bibnamefont  [1]{#1}%
\providecommand \bibfnamefont [1]{#1}%
\providecommand \citenamefont [1]{#1}%
\providecommand \href@noop [0]{\@secondoftwo}%
\providecommand \href [0]{\begingroup \@sanitize@url \@href}%
\providecommand \@href[1]{\@@startlink{#1}\@@href}%
\providecommand \@@href[1]{\endgroup#1\@@endlink}%
\providecommand \@sanitize@url [0]{\catcode `\\12\catcode `\$12\catcode
  `\&12\catcode `\#12\catcode `\^12\catcode `\_12\catcode `\%12\relax}%
\providecommand \@@startlink[1]{}%
\providecommand \@@endlink[0]{}%
\providecommand \url  [0]{\begingroup\@sanitize@url \@url }%
\providecommand \@url [1]{\endgroup\@href {#1}{\urlprefix }}%
\providecommand \urlprefix  [0]{URL }%
\providecommand \Eprint [0]{\href }%
\providecommand \doibase [0]{http://dx.doi.org/}%
\providecommand \selectlanguage [0]{\@gobble}%
\providecommand \bibinfo  [0]{\@secondoftwo}%
\providecommand \bibfield  [0]{\@secondoftwo}%
\providecommand \translation [1]{[#1]}%
\providecommand \BibitemOpen [0]{}%
\providecommand \bibitemStop [0]{}%
\providecommand \bibitemNoStop [0]{.\EOS\space}%
\providecommand \EOS [0]{\spacefactor3000\relax}%
\providecommand \BibitemShut  [1]{\csname bibitem#1\endcsname}%
\let\auto@bib@innerbib\@empty
\bibitem [{\citenamefont
  {Cavagna}(2009)}]{cavagnaSupercooledLiquidsPedestrians2009}%
  \BibitemOpen
  \bibfield  {author} {\bibinfo {author} {\bibfnamefont {A.}~\bibnamefont
  {Cavagna}},\ }\href {\doibase 10.1016/j.physrep.2009.03.003} {\bibfield
  {journal} {\bibinfo  {journal} {Phys. Rep.}\ }\textbf {\bibinfo {volume}
  {476}},\ \bibinfo {pages} {51} (\bibinfo {year} {2009})},\ \Eprint
  {http://arxiv.org/abs/0903.4264} {0903.4264} \BibitemShut {NoStop}%
\bibitem [{\citenamefont {Sastry}\ \emph {et~al.}(1998)\citenamefont {Sastry},
  \citenamefont {Debenedetti},\ and\ \citenamefont
  {Stillinger}}]{sastrySignaturesDistinctDynamical1998}%
  \BibitemOpen
  \bibfield  {author} {\bibinfo {author} {\bibfnamefont {S.}~\bibnamefont
  {Sastry}}, \bibinfo {author} {\bibfnamefont {P.~G.}\ \bibnamefont
  {Debenedetti}}, \ and\ \bibinfo {author} {\bibfnamefont {F.~H.}\ \bibnamefont
  {Stillinger}},\ }\href {\doibase 10.1038/31189} {\bibfield  {journal}
  {\bibinfo  {journal} {Nature}\ }\textbf {\bibinfo {volume} {393}},\ \bibinfo
  {pages} {554} (\bibinfo {year} {1998})}\BibitemShut {NoStop}%
\bibitem [{\citenamefont {Berthier}\ and\ \citenamefont
  {Biroli}(2011)}]{berthierTheoreticalPerspectiveGlass2011}%
  \BibitemOpen
  \bibfield  {author} {\bibinfo {author} {\bibfnamefont {L.}~\bibnamefont
  {Berthier}}\ and\ \bibinfo {author} {\bibfnamefont {G.}~\bibnamefont
  {Biroli}},\ }\href {\doibase 10.1103/RevModPhys.83.587} {\bibfield  {journal}
  {\bibinfo  {journal} {Rev. Mod. Phys.}\ }\textbf {\bibinfo {volume} {83}},\
  \bibinfo {pages} {587} (\bibinfo {year} {2011})}\BibitemShut {NoStop}%
\bibitem [{\citenamefont {Young}(1998)}]{youngSpinGlassesRandom1998}%
  \BibitemOpen
  \bibfield  {author} {\bibinfo {author} {\bibfnamefont {A.~P.}\ \bibnamefont
  {Young}},\ }\href@noop {} {\emph {\bibinfo {title} {Spin {{Glasses}} and
  {{Random Fields}}}}}\ (\bibinfo  {publisher} {World Scientific},\ \bibinfo
  {year} {1998})\BibitemShut {NoStop}%
\bibitem [{\citenamefont {Nattermann}\ and\ \citenamefont
  {Scheidl}(2000)}]{nattermannVortexglassPhasesTypeII2000}%
  \BibitemOpen
  \bibfield  {author} {\bibinfo {author} {\bibfnamefont {T.}~\bibnamefont
  {Nattermann}}\ and\ \bibinfo {author} {\bibfnamefont {S.}~\bibnamefont
  {Scheidl}},\ }\href {\doibase 10.1080/000187300412257} {\bibfield  {journal}
  {\bibinfo  {journal} {Adv. Phys.}\ }\textbf {\bibinfo {volume} {49}},\
  \bibinfo {pages} {607} (\bibinfo {year} {2000})}\BibitemShut {NoStop}%
\bibitem [{\citenamefont
  {Tanaka}(1999{\natexlab{a}})}]{tanakaTwoorderparameterDescriptionLiquids1999}%
  \BibitemOpen
  \bibfield  {author} {\bibinfo {author} {\bibfnamefont {H.}~\bibnamefont
  {Tanaka}},\ }\href {\doibase 10.1063/1.479596} {\bibfield  {journal}
  {\bibinfo  {journal} {J. Chem. Phys.}\ }\textbf {\bibinfo {volume} {111}},\
  \bibinfo {pages} {3163} (\bibinfo {year} {1999}{\natexlab{a}})}\BibitemShut
  {NoStop}%
\bibitem [{\citenamefont {Tarjus}\ \emph {et~al.}(2005)\citenamefont {Tarjus},
  \citenamefont {Kivelson}, \citenamefont {Nussinov},\ and\ \citenamefont
  {Viot}}]{tarjusfrustrationbasedapproachsupercooled2005}%
  \BibitemOpen
  \bibfield  {author} {\bibinfo {author} {\bibfnamefont {G.}~\bibnamefont
  {Tarjus}}, \bibinfo {author} {\bibfnamefont {S.~A.}\ \bibnamefont
  {Kivelson}}, \bibinfo {author} {\bibfnamefont {Z.}~\bibnamefont {Nussinov}},
  \ and\ \bibinfo {author} {\bibfnamefont {P.}~\bibnamefont {Viot}},\ }\href
  {\doibase 10.1088/0953-8984/17/50/R01} {\bibfield  {journal} {\bibinfo
  {journal} {J. Phys.: Condens. Matter}\ }\textbf {\bibinfo {volume} {17}},\
  \bibinfo {pages} {R1143} (\bibinfo {year} {2005})}\BibitemShut {NoStop}%
\bibitem [{\citenamefont {Parisi}\ \emph {et~al.}(2020)\citenamefont {Parisi},
  \citenamefont {Urbani},\ and\ \citenamefont
  {Zamponi}}]{parisiTheorySimpleGlasses2020}%
  \BibitemOpen
  \bibfield  {author} {\bibinfo {author} {\bibfnamefont {G.}~\bibnamefont
  {Parisi}}, \bibinfo {author} {\bibfnamefont {P.}~\bibnamefont {Urbani}}, \
  and\ \bibinfo {author} {\bibfnamefont {F.}~\bibnamefont {Zamponi}},\
  }\href@noop {} {\emph {\bibinfo {title} {Theory of {{Simple Glasses}}:
  {{Exact Solutions}} in {{Infinite Dimensions}}}}}\ (\bibinfo  {publisher}
  {Cambridge University Press},\ \bibinfo {address} {New York},\ \bibinfo
  {year} {2020})\BibitemShut {NoStop}%
\bibitem [{\citenamefont {Brumer}\ and\ \citenamefont
  {Reichman}(2004)}]{brumerMeanfieldTheoryModecoupling2004}%
  \BibitemOpen
  \bibfield  {author} {\bibinfo {author} {\bibfnamefont {Y.}~\bibnamefont
  {Brumer}}\ and\ \bibinfo {author} {\bibfnamefont {D.~R.}\ \bibnamefont
  {Reichman}},\ }\href {\doibase 10.1103/PhysRevE.69.041202} {\bibfield
  {journal} {\bibinfo  {journal} {Phys. Rev. E}\ }\textbf {\bibinfo {volume}
  {69}},\ \bibinfo {pages} {041202} (\bibinfo {year} {2004})}\BibitemShut
  {NoStop}%
\bibitem [{\citenamefont {Biroli}\ and\ \citenamefont
  {Bouchaud}(2012)}]{biroliRandomFirstOrderTransition2012}%
  \BibitemOpen
  \bibfield  {author} {\bibinfo {author} {\bibfnamefont {G.}~\bibnamefont
  {Biroli}}\ and\ \bibinfo {author} {\bibfnamefont {J.-P.}\ \bibnamefont
  {Bouchaud}},\ }in\ \href {\doibase 10.1002/9781118202470.ch2} {\emph
  {\bibinfo {booktitle} {Structural {{Glasses}} and {{Supercooled Liquids}}}}}\
  (\bibinfo  {publisher} {John Wiley \& Sons, Ltd},\ \bibinfo {year} {2012})\
  Chap.~\bibinfo {chapter} {2}, pp.\ \bibinfo {pages} {31--113}\BibitemShut
  {NoStop}%
\bibitem [{\citenamefont {Horbach}\ and\ \citenamefont
  {Kob}(1999)}]{horbachStaticDynamicProperties1999}%
  \BibitemOpen
  \bibfield  {author} {\bibinfo {author} {\bibfnamefont {J.}~\bibnamefont
  {Horbach}}\ and\ \bibinfo {author} {\bibfnamefont {W.}~\bibnamefont {Kob}},\
  }\href {\doibase 10.1103/PhysRevB.60.3169} {\bibfield  {journal} {\bibinfo
  {journal} {Phys. Rev. B}\ }\textbf {\bibinfo {volume} {60}},\ \bibinfo
  {pages} {3169} (\bibinfo {year} {1999})}\BibitemShut {NoStop}%
\bibitem [{\citenamefont {Greet}\ and\ \citenamefont
  {Magill}(1967)}]{greetEmpiricalCorrespondingstatesRelationship1967}%
  \BibitemOpen
  \bibfield  {author} {\bibinfo {author} {\bibfnamefont {R.~J.}\ \bibnamefont
  {Greet}}\ and\ \bibinfo {author} {\bibfnamefont {J.~H.}\ \bibnamefont
  {Magill}},\ }\href {\doibase 10.1021/j100865a030} {\bibfield  {journal}
  {\bibinfo  {journal} {J. Phys. Chem.}\ }\textbf {\bibinfo {volume} {71}},\
  \bibinfo {pages} {1746} (\bibinfo {year} {1967})}\BibitemShut {NoStop}%
\bibitem [{\citenamefont
  {Tanaka}(1999{\natexlab{b}})}]{tanakaTwoorderparameterDescriptionLiquids1999a}%
  \BibitemOpen
  \bibfield  {author} {\bibinfo {author} {\bibfnamefont {H.}~\bibnamefont
  {Tanaka}},\ }\href {\doibase 10.1063/1.479597} {\bibfield  {journal}
  {\bibinfo  {journal} {J. Chem. Phys.}\ }\textbf {\bibinfo {volume} {111}},\
  \bibinfo {pages} {3175} (\bibinfo {year} {1999}{\natexlab{b}})}\BibitemShut
  {NoStop}%
\bibitem [{\citenamefont {Hansen}\ \emph {et~al.}(1997)\citenamefont {Hansen},
  \citenamefont {Stickel}, \citenamefont {Berger}, \citenamefont {Richert},\
  and\ \citenamefont {Fischer}}]{hansenDynamicsGlassformingLiquids1997}%
  \BibitemOpen
  \bibfield  {author} {\bibinfo {author} {\bibfnamefont {C.}~\bibnamefont
  {Hansen}}, \bibinfo {author} {\bibfnamefont {F.}~\bibnamefont {Stickel}},
  \bibinfo {author} {\bibfnamefont {T.}~\bibnamefont {Berger}}, \bibinfo
  {author} {\bibfnamefont {R.}~\bibnamefont {Richert}}, \ and\ \bibinfo
  {author} {\bibfnamefont {E.~W.}\ \bibnamefont {Fischer}},\ }\href {\doibase
  doi:10.1063/1.474456} {\bibfield  {journal} {\bibinfo  {journal} {J. Chem.
  Phys.}\ }\textbf {\bibinfo {volume} {107}},\ \bibinfo {pages} {1086}
  (\bibinfo {year} {1997})}\BibitemShut {NoStop}%
\bibitem [{\citenamefont {Banerjee}\ \emph {et~al.}(2017)\citenamefont
  {Banerjee}, \citenamefont {Nandi}, \citenamefont {Sastry},\ and\
  \citenamefont
  {Maitra~Bhattacharyya}}]{banerjeeDeterminationOnsetTemperature2017}%
  \BibitemOpen
  \bibfield  {author} {\bibinfo {author} {\bibfnamefont {A.}~\bibnamefont
  {Banerjee}}, \bibinfo {author} {\bibfnamefont {M.~K.}\ \bibnamefont {Nandi}},
  \bibinfo {author} {\bibfnamefont {S.}~\bibnamefont {Sastry}}, \ and\ \bibinfo
  {author} {\bibfnamefont {S.}~\bibnamefont {Maitra~Bhattacharyya}},\ }\href
  {\doibase 10.1063/1.4991848} {\bibfield  {journal} {\bibinfo  {journal} {J.
  Chem. Phys.}\ }\textbf {\bibinfo {volume} {147}},\ \bibinfo {pages} {024504}
  (\bibinfo {year} {2017})}\BibitemShut {NoStop}%
\bibitem [{\citenamefont {Pedersen}\ \emph {et~al.}(2018)\citenamefont
  {Pedersen}, \citenamefont {Schr{\o}der},\ and\ \citenamefont
  {Dyre}}]{pedersenPhaseDiagramKobAndersenType2018}%
  \BibitemOpen
  \bibfield  {author} {\bibinfo {author} {\bibfnamefont {U.~R.}\ \bibnamefont
  {Pedersen}}, \bibinfo {author} {\bibfnamefont {T.~B.}\ \bibnamefont
  {Schr{\o}der}}, \ and\ \bibinfo {author} {\bibfnamefont {J.~C.}\ \bibnamefont
  {Dyre}},\ }\href {\doibase 10.1103/PhysRevLett.120.165501} {\bibfield
  {journal} {\bibinfo  {journal} {Phys. Rev. Lett.}\ }\textbf {\bibinfo
  {volume} {120}},\ \bibinfo {pages} {165501} (\bibinfo {year}
  {2018})}\BibitemShut {NoStop}%
\bibitem [{\citenamefont {Bell}\ \emph {et~al.}(2020)\citenamefont {Bell},
  \citenamefont {Dyre},\ and\ \citenamefont
  {Ingebrigtsen}}]{bellExcessentropyScalingSupercooled2020}%
  \BibitemOpen
  \bibfield  {author} {\bibinfo {author} {\bibfnamefont {I.~H.}\ \bibnamefont
  {Bell}}, \bibinfo {author} {\bibfnamefont {J.~C.}\ \bibnamefont {Dyre}}, \
  and\ \bibinfo {author} {\bibfnamefont {T.~S.}\ \bibnamefont {Ingebrigtsen}},\
  }\href {\doibase 10.1038/s41467-020-17948-1} {\bibfield  {journal} {\bibinfo
  {journal} {Nat. Commun.}\ }\textbf {\bibinfo {volume} {11}},\ \bibinfo
  {pages} {4300} (\bibinfo {year} {2020})}\BibitemShut {NoStop}%
\bibitem [{\citenamefont {Coslovich}\ and\ \citenamefont
  {Pastore}(2007)}]{coslovichUnderstandingFragilitySupercooled2007b}%
  \BibitemOpen
  \bibfield  {author} {\bibinfo {author} {\bibfnamefont {D.}~\bibnamefont
  {Coslovich}}\ and\ \bibinfo {author} {\bibfnamefont {G.}~\bibnamefont
  {Pastore}},\ }\href {\doibase 10.1063/1.2773720} {\bibfield  {journal}
  {\bibinfo  {journal} {J. Chem. Phys.}\ }\textbf {\bibinfo {volume} {127}},\
  \bibinfo {pages} {124505} (\bibinfo {year} {2007})}\BibitemShut {NoStop}%
\bibitem [{\citenamefont {Schoenholz}\ \emph {et~al.}(2016)\citenamefont
  {Schoenholz}, \citenamefont {Cubuk}, \citenamefont {Sussman}, \citenamefont
  {Kaxiras},\ and\ \citenamefont
  {Liu}}]{schoenholzStructuralApproachRelaxation2016}%
  \BibitemOpen
  \bibfield  {author} {\bibinfo {author} {\bibfnamefont {S.~S.}\ \bibnamefont
  {Schoenholz}}, \bibinfo {author} {\bibfnamefont {E.~D.}\ \bibnamefont
  {Cubuk}}, \bibinfo {author} {\bibfnamefont {D.~M.}\ \bibnamefont {Sussman}},
  \bibinfo {author} {\bibfnamefont {E.}~\bibnamefont {Kaxiras}}, \ and\
  \bibinfo {author} {\bibfnamefont {A.~J.}\ \bibnamefont {Liu}},\ }\href
  {\doibase 10.1038/nphys3644} {\bibfield  {journal} {\bibinfo  {journal} {Nat.
  Phys.}\ }\textbf {\bibinfo {volume} {12}},\ \bibinfo {pages} {469} (\bibinfo
  {year} {2016})}\BibitemShut {NoStop}%
\bibitem [{\citenamefont {Singh}\ \emph {et~al.}(2021)\citenamefont {Singh},
  \citenamefont {Bhattacharyya},\ and\ \citenamefont
  {Singh}}]{singhEmergenceCooperativelyReorganizing2021}%
  \BibitemOpen
  \bibfield  {author} {\bibinfo {author} {\bibfnamefont {A.}~\bibnamefont
  {Singh}}, \bibinfo {author} {\bibfnamefont {S.~M.}\ \bibnamefont
  {Bhattacharyya}}, \ and\ \bibinfo {author} {\bibfnamefont {Y.}~\bibnamefont
  {Singh}},\ }\href {\doibase 10.1103/PhysRevE.103.032611} {\bibfield
  {journal} {\bibinfo  {journal} {Phys. Rev. E}\ }\textbf {\bibinfo {volume}
  {103}},\ \bibinfo {pages} {032611} (\bibinfo {year} {2021})}\BibitemShut
  {NoStop}%
\bibitem [{\citenamefont {Kob}\ and\ \citenamefont
  {Andersen}(1995)}]{kobTestingModecouplingTheory1995a}%
  \BibitemOpen
  \bibfield  {author} {\bibinfo {author} {\bibfnamefont {W.}~\bibnamefont
  {Kob}}\ and\ \bibinfo {author} {\bibfnamefont {H.~C.}\ \bibnamefont
  {Andersen}},\ }\href {\doibase 10.1103/PhysRevE.52.4134} {\bibfield
  {journal} {\bibinfo  {journal} {Phys. Rev. E}\ }\textbf {\bibinfo {volume}
  {52}},\ \bibinfo {pages} {4134} (\bibinfo {year} {1995})}\BibitemShut
  {NoStop}%
\bibitem [{\citenamefont {Coslovich}(2025)}]{atooms}%
  \BibitemOpen
  \bibfield  {author} {\bibinfo {author} {\bibfnamefont {D.}~\bibnamefont
  {Coslovich}},\ }\href {\doibase 10.5281/zenodo.1183301} {\enquote {\bibinfo
  {title} {atooms: A framework for simulations of interacting particles
  (3.22.0)}}\ } (\bibinfo {year} {2025})\BibitemShut {NoStop}%
\bibitem [{\citenamefont {Bailey}\ \emph {et~al.}(2017)\citenamefont {Bailey},
  \citenamefont {Hansen}, \citenamefont {Ingebrigtsen}, \citenamefont
  {Veldhorst}, \citenamefont {B{\o}hling}, \citenamefont {Lemarchand},
  \citenamefont {Olsen}, \citenamefont {Bacher}, \citenamefont {Costigliola},
  \citenamefont {Pedersen}, \citenamefont {Larsen}, \citenamefont {Dyre},\ and\
  \citenamefont {Schr{\o}der}}]{baileyRUMDGeneralPurpose2017}%
  \BibitemOpen
  \bibfield  {author} {\bibinfo {author} {\bibfnamefont {N.}~\bibnamefont
  {Bailey}}, \bibinfo {author} {\bibfnamefont {J.~S.}\ \bibnamefont {Hansen}},
  \bibinfo {author} {\bibfnamefont {T.}~\bibnamefont {Ingebrigtsen}}, \bibinfo
  {author} {\bibfnamefont {A.}~\bibnamefont {Veldhorst}}, \bibinfo {author}
  {\bibfnamefont {L.}~\bibnamefont {B{\o}hling}}, \bibinfo {author}
  {\bibfnamefont {C.}~\bibnamefont {Lemarchand}}, \bibinfo {author}
  {\bibfnamefont {A.}~\bibnamefont {Olsen}}, \bibinfo {author} {\bibfnamefont
  {A.}~\bibnamefont {Bacher}}, \bibinfo {author} {\bibfnamefont
  {L.}~\bibnamefont {Costigliola}}, \bibinfo {author} {\bibfnamefont
  {U.}~\bibnamefont {Pedersen}}, \bibinfo {author} {\bibfnamefont
  {H.}~\bibnamefont {Larsen}}, \bibinfo {author} {\bibfnamefont
  {J.}~\bibnamefont {Dyre}}, \ and\ \bibinfo {author} {\bibfnamefont
  {T.}~\bibnamefont {Schr{\o}der}},\ }\href {\doibase
  10.21468/SciPostPhys.3.6.038} {\bibfield  {journal} {\bibinfo  {journal}
  {SciPost Phys.}\ }\textbf {\bibinfo {volume} {3}},\ \bibinfo {pages} {038}
  (\bibinfo {year} {2017})}\BibitemShut {NoStop}%
\bibitem [{si()}]{si}%
  \BibitemOpen
  \href@noop {} {}\bibinfo {note} {See Supplementary Material at \url{https://arxiv.org/src/2406.04921v2/anc/supplement.pdf}
  for simulation details, additional analysis and tabulated data.}\BibitemShut
  {Stop}%
\bibitem [{\citenamefont {Brazhkin}\ \emph {et~al.}(2013)\citenamefont
  {Brazhkin}, \citenamefont {Fomin}, \citenamefont {Lyapin}, \citenamefont
  {Ryzhov}, \citenamefont {Tsiok},\ and\ \citenamefont
  {Trachenko}}]{Brazhkin_Fomin_Lyapin_Ryzhov_Tsiok_Trachenko_2013}%
  \BibitemOpen
  \bibfield  {author} {\bibinfo {author} {\bibfnamefont {V.~V.}\ \bibnamefont
  {Brazhkin}}, \bibinfo {author} {\bibfnamefont {Y.~D.}\ \bibnamefont {Fomin}},
  \bibinfo {author} {\bibfnamefont {A.~G.}\ \bibnamefont {Lyapin}}, \bibinfo
  {author} {\bibfnamefont {V.~N.}\ \bibnamefont {Ryzhov}}, \bibinfo {author}
  {\bibfnamefont {E.~N.}\ \bibnamefont {Tsiok}}, \ and\ \bibinfo {author}
  {\bibfnamefont {K.}~\bibnamefont {Trachenko}},\ }\href {\doibase
  10.1103/PhysRevLett.111.145901} {\bibfield  {journal} {\bibinfo  {journal}
  {Phys. Rev. Lett.}\ }\textbf {\bibinfo {volume} {111}},\ \bibinfo {pages}
  {145901} (\bibinfo {year} {2013})}\BibitemShut {NoStop}%
\bibitem [{\citenamefont {Costigliola}\ \emph {et~al.}(2018)\citenamefont
  {Costigliola}, \citenamefont {Pedersen}, \citenamefont {Heyes}, \citenamefont
  {Schr{\o}der},\ and\ \citenamefont
  {Dyre}}]{costigliolaCommunicationSimpleLiquids2018}%
  \BibitemOpen
  \bibfield  {author} {\bibinfo {author} {\bibfnamefont {L.}~\bibnamefont
  {Costigliola}}, \bibinfo {author} {\bibfnamefont {U.~R.}\ \bibnamefont
  {Pedersen}}, \bibinfo {author} {\bibfnamefont {D.~M.}\ \bibnamefont {Heyes}},
  \bibinfo {author} {\bibfnamefont {T.~B.}\ \bibnamefont {Schr{\o}der}}, \ and\
  \bibinfo {author} {\bibfnamefont {J.~C.}\ \bibnamefont {Dyre}},\ }\href
  {\doibase 10.1063/1.5022058} {\bibfield  {journal} {\bibinfo  {journal} {J.
  Chem. Phys.}\ }\textbf {\bibinfo {volume} {148}},\ \bibinfo {pages} {081101}
  (\bibinfo {year} {2018})}\BibitemShut {NoStop}%
\bibitem [{Note1()}]{Note1}%
  \BibitemOpen
  \bibinfo {note} {In Ref.~\protect \rev@citealp
  {pedersenPhaseDiagramKobAndersenType2018} the freezing line was referred to
  as ``melting line''.}\BibitemShut {Stop}%
\bibitem [{\citenamefont {Silbey}\ \emph {et~al.}(2004)\citenamefont {Silbey},
  \citenamefont {Alberty},\ and\ \citenamefont
  {Bawendi}}]{silbeyPhysicalChemistry2004}%
  \BibitemOpen
  \bibfield  {author} {\bibinfo {author} {\bibfnamefont {R.~J.}\ \bibnamefont
  {Silbey}}, \bibinfo {author} {\bibfnamefont {R.~A.}\ \bibnamefont {Alberty}},
  \ and\ \bibinfo {author} {\bibfnamefont {M.~G.}\ \bibnamefont {Bawendi}},\
  }\href@noop {} {\emph {\bibinfo {title} {Physical {{Chemistry}}}}},\ \bibinfo
  {edition} {4th}\ ed.\ (\bibinfo  {publisher} {Wiley},\ \bibinfo {address}
  {Hoboken, NJ},\ \bibinfo {year} {2004})\BibitemShut {NoStop}%
\bibitem [{\citenamefont
  {Dyre}(2018)}]{dyrePerspectiveExcessentropyScaling2018}%
  \BibitemOpen
  \bibfield  {author} {\bibinfo {author} {\bibfnamefont {J.~C.}\ \bibnamefont
  {Dyre}},\ }\href {\doibase 10.1063/1.5055064} {\bibfield  {journal} {\bibinfo
   {journal} {J. Chem. Phys.}\ }\textbf {\bibinfo {volume} {149}},\ \bibinfo
  {pages} {210901} (\bibinfo {year} {2018})}\BibitemShut {NoStop}%
\bibitem [{\citenamefont {Schr{\o}der}\ \emph {et~al.}(2011)\citenamefont
  {Schr{\o}der}, \citenamefont {Gnan}, \citenamefont {Pedersen}, \citenamefont
  {Bailey},\ and\ \citenamefont
  {Dyre}}]{schroderPressureenergyCorrelationsLiquids2011}%
  \BibitemOpen
  \bibfield  {author} {\bibinfo {author} {\bibfnamefont {T.~B.}\ \bibnamefont
  {Schr{\o}der}}, \bibinfo {author} {\bibfnamefont {N.}~\bibnamefont {Gnan}},
  \bibinfo {author} {\bibfnamefont {U.~R.}\ \bibnamefont {Pedersen}}, \bibinfo
  {author} {\bibfnamefont {N.~P.}\ \bibnamefont {Bailey}}, \ and\ \bibinfo
  {author} {\bibfnamefont {J.~C.}\ \bibnamefont {Dyre}},\ }\href {\doibase
  10.1063/1.3582900} {\bibfield  {journal} {\bibinfo  {journal} {J. Chem.
  Phys.}\ }\textbf {\bibinfo {volume} {134}},\ \bibinfo {pages} {164505}
  (\bibinfo {year} {2011})}\BibitemShut {NoStop}%
\bibitem [{\citenamefont {Ingebrigtsen}\ \emph {et~al.}(2012)\citenamefont
  {Ingebrigtsen}, \citenamefont {B{\o}hling}, \citenamefont {Schr{\o}der},\
  and\ \citenamefont
  {Dyre}}]{ingebrigtsenCommunicationThermodynamicsCondensed2012}%
  \BibitemOpen
  \bibfield  {author} {\bibinfo {author} {\bibfnamefont {T.~S.}\ \bibnamefont
  {Ingebrigtsen}}, \bibinfo {author} {\bibfnamefont {L.}~\bibnamefont
  {B{\o}hling}}, \bibinfo {author} {\bibfnamefont {T.~B.}\ \bibnamefont
  {Schr{\o}der}}, \ and\ \bibinfo {author} {\bibfnamefont {J.~C.}\ \bibnamefont
  {Dyre}},\ }\href {\doibase doi:10.1063/1.3685804} {\bibfield  {journal}
  {\bibinfo  {journal} {J. Chem. Phys.}\ }\textbf {\bibinfo {volume} {136}},\
  \bibinfo {pages} {061102} (\bibinfo {year} {2012})}\BibitemShut {NoStop}%
\bibitem [{\citenamefont {Pedersen}\ \emph {et~al.}(2016)\citenamefont
  {Pedersen}, \citenamefont {Costigliola}, \citenamefont {Bailey},
  \citenamefont {Schr{\o}der},\ and\ \citenamefont
  {Dyre}}]{pedersenThermodynamicsFreezingMelting2016}%
  \BibitemOpen
  \bibfield  {author} {\bibinfo {author} {\bibfnamefont {U.~R.}\ \bibnamefont
  {Pedersen}}, \bibinfo {author} {\bibfnamefont {L.}~\bibnamefont
  {Costigliola}}, \bibinfo {author} {\bibfnamefont {N.~P.}\ \bibnamefont
  {Bailey}}, \bibinfo {author} {\bibfnamefont {T.~B.}\ \bibnamefont
  {Schr{\o}der}}, \ and\ \bibinfo {author} {\bibfnamefont {J.~C.}\ \bibnamefont
  {Dyre}},\ }\href {\doibase 10.1038/ncomms12386} {\bibfield  {journal}
  {\bibinfo  {journal} {Nat. Commun.}\ }\textbf {\bibinfo {volume} {7}},\
  \bibinfo {pages} {12386} (\bibinfo {year} {2016})}\BibitemShut {NoStop}%
\bibitem [{\citenamefont {Singh}\ \emph {et~al.}(2012)\citenamefont {Singh},
  \citenamefont {Agarwal}, \citenamefont {Dhabal},\ and\ \citenamefont
  {Chakravarty}}]{singhStructuralCorrelationsCooperative2012a}%
  \BibitemOpen
  \bibfield  {author} {\bibinfo {author} {\bibfnamefont {M.}~\bibnamefont
  {Singh}}, \bibinfo {author} {\bibfnamefont {M.}~\bibnamefont {Agarwal}},
  \bibinfo {author} {\bibfnamefont {D.}~\bibnamefont {Dhabal}}, \ and\ \bibinfo
  {author} {\bibfnamefont {C.}~\bibnamefont {Chakravarty}},\ }\href {\doibase
  10.1063/1.4731705} {\bibfield  {journal} {\bibinfo  {journal} {J. Chem.
  Phys.}\ }\textbf {\bibinfo {volume} {137}},\ \bibinfo {pages} {024508}
  (\bibinfo {year} {2012})}\BibitemShut {NoStop}%
\bibitem [{\citenamefont {Baranyai}\ and\ \citenamefont
  {Evans}(1989)}]{baranyaiDirectEntropyCalculation1989}%
  \BibitemOpen
  \bibfield  {author} {\bibinfo {author} {\bibfnamefont {A.}~\bibnamefont
  {Baranyai}}\ and\ \bibinfo {author} {\bibfnamefont {D.~J.}\ \bibnamefont
  {Evans}},\ }\href {\doibase 10.1103/PhysRevA.40.3817} {\bibfield  {journal}
  {\bibinfo  {journal} {Phys. Rev. A}\ }\textbf {\bibinfo {volume} {40}},\
  \bibinfo {pages} {3817} (\bibinfo {year} {1989})}\BibitemShut {NoStop}%
\bibitem [{\citenamefont {Giaquinta}\ \emph {et~al.}(1992)\citenamefont
  {Giaquinta}, \citenamefont {Giunta},\ and\ \citenamefont
  {Prestipino~Giarritta}}]{giaquintaEntropyFreezingSimple1992}%
  \BibitemOpen
  \bibfield  {author} {\bibinfo {author} {\bibfnamefont {P.~V.}\ \bibnamefont
  {Giaquinta}}, \bibinfo {author} {\bibfnamefont {G.}~\bibnamefont {Giunta}}, \
  and\ \bibinfo {author} {\bibfnamefont {S.}~\bibnamefont
  {Prestipino~Giarritta}},\ }\href {\doibase 10.1103/PhysRevA.45.R6966}
  {\bibfield  {journal} {\bibinfo  {journal} {Phys. Rev. A}\ }\textbf {\bibinfo
  {volume} {45}},\ \bibinfo {pages} {R6966} (\bibinfo {year}
  {1992})}\BibitemShut {NoStop}%
\bibitem [{\citenamefont {Krekelberg}\ \emph {et~al.}(2008)\citenamefont
  {Krekelberg}, \citenamefont {Shen}, \citenamefont {Errington},\ and\
  \citenamefont {Truskett}}]{krekelbergResidualMultiparticleEntropy2008a}%
  \BibitemOpen
  \bibfield  {author} {\bibinfo {author} {\bibfnamefont {W.~P.}\ \bibnamefont
  {Krekelberg}}, \bibinfo {author} {\bibfnamefont {V.~K.}\ \bibnamefont
  {Shen}}, \bibinfo {author} {\bibfnamefont {J.~R.}\ \bibnamefont {Errington}},
  \ and\ \bibinfo {author} {\bibfnamefont {T.~M.}\ \bibnamefont {Truskett}},\
  }\href {\doibase 10.1063/1.2916697} {\bibfield  {journal} {\bibinfo
  {journal} {J. Chem. Phys.}\ }\textbf {\bibinfo {volume} {128}},\ \bibinfo
  {pages} {161101} (\bibinfo {year} {2008})}\BibitemShut {NoStop}%
\bibitem [{\citenamefont {Yiannourakou}\ \emph {et~al.}(2009)\citenamefont
  {Yiannourakou}, \citenamefont {Economou},\ and\ \citenamefont
  {Bitsanis}}]{yiannourakouPhaseEquilibriumColloidal2009a}%
  \BibitemOpen
  \bibfield  {author} {\bibinfo {author} {\bibfnamefont {M.}~\bibnamefont
  {Yiannourakou}}, \bibinfo {author} {\bibfnamefont {I.~G.}\ \bibnamefont
  {Economou}}, \ and\ \bibinfo {author} {\bibfnamefont {I.~A.}\ \bibnamefont
  {Bitsanis}},\ }\href {\doibase 10.1063/1.3131691} {\bibfield  {journal}
  {\bibinfo  {journal} {J. Chem. Phys.}\ }\textbf {\bibinfo {volume} {130}},\
  \bibinfo {pages} {194902} (\bibinfo {year} {2009})}\BibitemShut {NoStop}%
\bibitem [{\citenamefont {Russo}\ \emph {et~al.}(2018)\citenamefont {Russo},
  \citenamefont {Romano},\ and\ \citenamefont
  {Tanaka}}]{russoGlassFormingAbility2018}%
  \BibitemOpen
  \bibfield  {author} {\bibinfo {author} {\bibfnamefont {J.}~\bibnamefont
  {Russo}}, \bibinfo {author} {\bibfnamefont {F.}~\bibnamefont {Romano}}, \
  and\ \bibinfo {author} {\bibfnamefont {H.}~\bibnamefont {Tanaka}},\ }\href
  {\doibase 10.1103/PhysRevX.8.021040} {\bibfield  {journal} {\bibinfo
  {journal} {Phys. Rev. X}\ }\textbf {\bibinfo {volume} {8}},\ \bibinfo {pages}
  {021040} (\bibinfo {year} {2018})}\BibitemShut {NoStop}%
\bibitem [{\citenamefont {Molinero}\ \emph {et~al.}(2006)\citenamefont
  {Molinero}, \citenamefont {Sastry},\ and\ \citenamefont
  {Angell}}]{molineroTuningTetrahedralitySilicon2006}%
  \BibitemOpen
  \bibfield  {author} {\bibinfo {author} {\bibfnamefont {V.}~\bibnamefont
  {Molinero}}, \bibinfo {author} {\bibfnamefont {S.}~\bibnamefont {Sastry}}, \
  and\ \bibinfo {author} {\bibfnamefont {C.~A.}\ \bibnamefont {Angell}},\
  }\href {\doibase 10.1103/PhysRevLett.97.075701} {\bibfield  {journal}
  {\bibinfo  {journal} {Phys. Rev. Lett.}\ }\textbf {\bibinfo {volume} {97}},\
  \bibinfo {pages} {075701} (\bibinfo {year} {2006})}\BibitemShut {NoStop}%
\bibitem [{\citenamefont {Nishio}(2024)}]{nishioLiquidusCurveLennardJones2024}%
  \BibitemOpen
  \bibfield  {author} {\bibinfo {author} {\bibfnamefont {K.}~\bibnamefont
  {Nishio}},\ }\href {\doibase 10.1103/PhysRevE.109.044110} {\bibfield
  {journal} {\bibinfo  {journal} {Phys. Rev. E}\ }\textbf {\bibinfo {volume}
  {109}},\ \bibinfo {pages} {044110} (\bibinfo {year} {2024})}\BibitemShut
  {NoStop}%
\bibitem [{\citenamefont {Blochowicz}\ \emph {et~al.}(2003)\citenamefont
  {Blochowicz}, \citenamefont {Tschirwitz}, \citenamefont {Benkhof},\ and\
  \citenamefont {R{\"o}ssler}}]{blochowiczSusceptibilityFunctionsSlow2003}%
  \BibitemOpen
  \bibfield  {author} {\bibinfo {author} {\bibfnamefont {{\relax
  Th}.}~\bibnamefont {Blochowicz}}, \bibinfo {author} {\bibfnamefont {{\relax
  Ch}.}~\bibnamefont {Tschirwitz}}, \bibinfo {author} {\bibfnamefont {{\relax
  St}.}~\bibnamefont {Benkhof}}, \ and\ \bibinfo {author} {\bibfnamefont
  {E.~A.}\ \bibnamefont {R{\"o}ssler}},\ }\href {\doibase 10.1063/1.1563247}
  {\bibfield  {journal} {\bibinfo  {journal} {The Journal of Chemical Physics}\
  }\textbf {\bibinfo {volume} {118}},\ \bibinfo {pages} {7544} (\bibinfo {year}
  {2003})}\BibitemShut {NoStop}%
\bibitem [{\citenamefont {Schmidtke}\ \emph {et~al.}(2013)\citenamefont
  {Schmidtke}, \citenamefont {Petzold}, \citenamefont {Kahlau},\ and\
  \citenamefont {R{\"o}ssler}}]{schmidtkeReorientationalDynamicsMolecular2013}%
  \BibitemOpen
  \bibfield  {author} {\bibinfo {author} {\bibfnamefont {B.}~\bibnamefont
  {Schmidtke}}, \bibinfo {author} {\bibfnamefont {N.}~\bibnamefont {Petzold}},
  \bibinfo {author} {\bibfnamefont {R.}~\bibnamefont {Kahlau}}, \ and\ \bibinfo
  {author} {\bibfnamefont {E.~A.}\ \bibnamefont {R{\"o}ssler}},\ }\href
  {\doibase 10.1063/1.4817406} {\bibfield  {journal} {\bibinfo  {journal} {J.
  Chem. Phys.}\ }\textbf {\bibinfo {volume} {139}},\ \bibinfo {pages} {084504}
  (\bibinfo {year} {2013})}\BibitemShut {NoStop}%
\bibitem [{\citenamefont {Royall}\ and\ \citenamefont
  {Williams}(2015)}]{royallRoleLocalStructure2015}%
  \BibitemOpen
  \bibfield  {author} {\bibinfo {author} {\bibfnamefont {C.~P.}\ \bibnamefont
  {Royall}}\ and\ \bibinfo {author} {\bibfnamefont {S.~R.}\ \bibnamefont
  {Williams}},\ }\href {\doibase 10.1016/j.physrep.2014.11.004} {\bibfield
  {journal} {\bibinfo  {journal} {Phys. Rep.}\ }\textbf {\bibinfo {volume}
  {560}},\ \bibinfo {pages} {1} (\bibinfo {year} {2015})}\BibitemShut {NoStop}%
\bibitem [{\citenamefont {Tanaka}\ \emph {et~al.}(2019)\citenamefont {Tanaka},
  \citenamefont {Tong}, \citenamefont {Shi},\ and\ \citenamefont
  {Russo}}]{tanakaRevealingKeyStructural2019}%
  \BibitemOpen
  \bibfield  {author} {\bibinfo {author} {\bibfnamefont {H.}~\bibnamefont
  {Tanaka}}, \bibinfo {author} {\bibfnamefont {H.}~\bibnamefont {Tong}},
  \bibinfo {author} {\bibfnamefont {R.}~\bibnamefont {Shi}}, \ and\ \bibinfo
  {author} {\bibfnamefont {J.}~\bibnamefont {Russo}},\ }\href {\doibase
  10.1038/s42254-019-0053-3} {\bibfield  {journal} {\bibinfo  {journal} {Nat.
  Rev. Phys.}\ }\textbf {\bibinfo {volume} {1}},\ \bibinfo {pages} {333}
  (\bibinfo {year} {2019})}\BibitemShut {NoStop}%
\bibitem [{\citenamefont {Wei}\ \emph {et~al.}(2019)\citenamefont {Wei},
  \citenamefont {Yang}, \citenamefont {Jiang}, \citenamefont {Dai},
  \citenamefont {Wang}, \citenamefont {Dyre}, \citenamefont {Douglass},\ and\
  \citenamefont {Harrowell}}]{weiAssessingUtilityStructure2019}%
  \BibitemOpen
  \bibfield  {author} {\bibinfo {author} {\bibfnamefont {D.}~\bibnamefont
  {Wei}}, \bibinfo {author} {\bibfnamefont {J.}~\bibnamefont {Yang}}, \bibinfo
  {author} {\bibfnamefont {M.-Q.}\ \bibnamefont {Jiang}}, \bibinfo {author}
  {\bibfnamefont {L.-H.}\ \bibnamefont {Dai}}, \bibinfo {author} {\bibfnamefont
  {Y.-J.}\ \bibnamefont {Wang}}, \bibinfo {author} {\bibfnamefont {J.~C.}\
  \bibnamefont {Dyre}}, \bibinfo {author} {\bibfnamefont {I.}~\bibnamefont
  {Douglass}}, \ and\ \bibinfo {author} {\bibfnamefont {P.}~\bibnamefont
  {Harrowell}},\ }\href {\doibase 10.1063/1.5064531} {\bibfield  {journal}
  {\bibinfo  {journal} {J. Chem. Phys.}\ }\textbf {\bibinfo {volume} {150}},\
  \bibinfo {pages} {114502} (\bibinfo {year} {2019})}\BibitemShut {NoStop}%
\bibitem [{\citenamefont
  {Coslovich}(2011)}]{coslovichLocallyPreferredStructures2011}%
  \BibitemOpen
  \bibfield  {author} {\bibinfo {author} {\bibfnamefont {D.}~\bibnamefont
  {Coslovich}},\ }\href {\doibase 10.1103/PhysRevE.83.051505} {\bibfield
  {journal} {\bibinfo  {journal} {Phys. Rev. E}\ }\textbf {\bibinfo {volume}
  {83}},\ \bibinfo {pages} {051505} (\bibinfo {year} {2011})}\BibitemShut
  {NoStop}%
\bibitem [{\citenamefont {Crowther}\ \emph {et~al.}(2015)\citenamefont
  {Crowther}, \citenamefont {Turci},\ and\ \citenamefont
  {Royall}}]{crowtherNatureGeometricFrustration2015}%
  \BibitemOpen
  \bibfield  {author} {\bibinfo {author} {\bibfnamefont {P.}~\bibnamefont
  {Crowther}}, \bibinfo {author} {\bibfnamefont {F.}~\bibnamefont {Turci}}, \
  and\ \bibinfo {author} {\bibfnamefont {C.~P.}\ \bibnamefont {Royall}},\
  }\href {\doibase 10.1063/1.4927302} {\bibfield  {journal} {\bibinfo
  {journal} {J. Chem. Phys.}\ }\textbf {\bibinfo {volume} {143}},\ \bibinfo
  {pages} {044503} (\bibinfo {year} {2015})}\BibitemShut {NoStop}%
\bibitem [{\citenamefont {Nandi}\ \emph {et~al.}(2016)\citenamefont {Nandi},
  \citenamefont {Banerjee}, \citenamefont {Chakrabarty},\ and\ \citenamefont
  {Bhattacharyya}}]{nandiCompositionDependenceGlass2016a}%
  \BibitemOpen
  \bibfield  {author} {\bibinfo {author} {\bibfnamefont {U.~K.}\ \bibnamefont
  {Nandi}}, \bibinfo {author} {\bibfnamefont {A.}~\bibnamefont {Banerjee}},
  \bibinfo {author} {\bibfnamefont {S.}~\bibnamefont {Chakrabarty}}, \ and\
  \bibinfo {author} {\bibfnamefont {S.~M.}\ \bibnamefont {Bhattacharyya}},\
  }\href {\doibase 10.1063/1.4958630} {\bibfield  {journal} {\bibinfo
  {journal} {J. Chem. Phys.}\ }\textbf {\bibinfo {volume} {145}},\ \bibinfo
  {pages} {034503} (\bibinfo {year} {2016})}\BibitemShut {NoStop}%
\bibitem [{\citenamefont {Ingebrigtsen}\ \emph {et~al.}(2019)\citenamefont
  {Ingebrigtsen}, \citenamefont {Dyre}, \citenamefont {Schr{\o}der},\ and\
  \citenamefont
  {Royall}}]{ingebrigtsenCrystallizationInstabilityGlassForming2019}%
  \BibitemOpen
  \bibfield  {author} {\bibinfo {author} {\bibfnamefont {T.~S.}\ \bibnamefont
  {Ingebrigtsen}}, \bibinfo {author} {\bibfnamefont {J.~C.}\ \bibnamefont
  {Dyre}}, \bibinfo {author} {\bibfnamefont {T.~B.}\ \bibnamefont
  {Schr{\o}der}}, \ and\ \bibinfo {author} {\bibfnamefont {C.~P.}\ \bibnamefont
  {Royall}},\ }\href {\doibase 10.1103/PhysRevX.9.031016} {\bibfield  {journal}
  {\bibinfo  {journal} {Phys. Rev. X}\ }\textbf {\bibinfo {volume} {9}},\
  \bibinfo {pages} {031016} (\bibinfo {year} {2019})}\BibitemShut {NoStop}%
\bibitem [{\citenamefont {Pedersen}\ \emph {et~al.}(2021)\citenamefont
  {Pedersen}, \citenamefont {Douglass},\ and\ \citenamefont
  {Harrowell}}]{pedersenHowSupercooledLiquid2021}%
  \BibitemOpen
  \bibfield  {author} {\bibinfo {author} {\bibfnamefont {U.~R.}\ \bibnamefont
  {Pedersen}}, \bibinfo {author} {\bibfnamefont {I.}~\bibnamefont {Douglass}},
  \ and\ \bibinfo {author} {\bibfnamefont {P.}~\bibnamefont {Harrowell}},\
  }\href {\doibase 10.1063/5.0033206} {\bibfield  {journal} {\bibinfo
  {journal} {J. Chem. Phys.}\ }\textbf {\bibinfo {volume} {154}},\ \bibinfo
  {pages} {054503} (\bibinfo {year} {2021})}\BibitemShut {NoStop}%
\bibitem [{\citenamefont {Nauroth}\ and\ \citenamefont
  {Kob}(1997)}]{naurothQuantitativeTestModecoupling1997}%
  \BibitemOpen
  \bibfield  {author} {\bibinfo {author} {\bibfnamefont {M.}~\bibnamefont
  {Nauroth}}\ and\ \bibinfo {author} {\bibfnamefont {W.}~\bibnamefont {Kob}},\
  }\href {\doibase 10.1103/PhysRevE.55.657} {\bibfield  {journal} {\bibinfo
  {journal} {Phys. Rev. E}\ }\textbf {\bibinfo {volume} {55}},\ \bibinfo
  {pages} {657} (\bibinfo {year} {1997})}\BibitemShut {NoStop}%
\bibitem [{\citenamefont {Gotze}\ and\ \citenamefont
  {Voigtmann}(2003)}]{gotzeEffectCompositionChanges2003}%
  \BibitemOpen
  \bibfield  {author} {\bibinfo {author} {\bibfnamefont {W.}~\bibnamefont
  {Gotze}}\ and\ \bibinfo {author} {\bibfnamefont {{\relax Th}.}~\bibnamefont
  {Voigtmann}},\ }\href {\doibase 10.1103/PhysRevE.67.021502} {\bibfield
  {journal} {\bibinfo  {journal} {Phys. Rev. E}\ }\textbf {\bibinfo {volume}
  {67}},\ \bibinfo {pages} {021502} (\bibinfo {year} {2003})}\BibitemShut
  {NoStop}%
\bibitem [{\citenamefont {Luo}\ and\ \citenamefont
  {Janssen}(2020)}]{luoGeneralizedModecouplingTheory2020a}%
  \BibitemOpen
  \bibfield  {author} {\bibinfo {author} {\bibfnamefont {C.}~\bibnamefont
  {Luo}}\ and\ \bibinfo {author} {\bibfnamefont {L.~M.~C.}\ \bibnamefont
  {Janssen}},\ }\href {\doibase 10.1063/5.0026969} {\bibfield  {journal}
  {\bibinfo  {journal} {J. Chem. Phys.}\ }\textbf {\bibinfo {volume} {153}},\
  \bibinfo {pages} {214507} (\bibinfo {year} {2020})}\BibitemShut {NoStop}%
\bibitem [{\citenamefont {Ciarella}\ \emph {et~al.}(2021)\citenamefont
  {Ciarella}, \citenamefont {Luo}, \citenamefont {Debets},\ and\ \citenamefont
  {Janssen}}]{ciarellaMulticomponentGeneralizedModecoupling2021}%
  \BibitemOpen
  \bibfield  {author} {\bibinfo {author} {\bibfnamefont {S.}~\bibnamefont
  {Ciarella}}, \bibinfo {author} {\bibfnamefont {C.}~\bibnamefont {Luo}},
  \bibinfo {author} {\bibfnamefont {V.~E.}\ \bibnamefont {Debets}}, \ and\
  \bibinfo {author} {\bibfnamefont {L.~M.~C.}\ \bibnamefont {Janssen}},\ }\href
  {\doibase 10.1140/epje/s10189-021-00095-w} {\bibfield  {journal} {\bibinfo
  {journal} {Eur. Phys. J. E}\ }\textbf {\bibinfo {volume} {44}},\ \bibinfo
  {pages} {91} (\bibinfo {year} {2021})}\BibitemShut {NoStop}%
\bibitem [{\citenamefont {Sun}\ and\ \citenamefont
  {Harrowell}(2022)}]{sunGeneralStructuralOrder2022}%
  \BibitemOpen
  \bibfield  {author} {\bibinfo {author} {\bibfnamefont {G.}~\bibnamefont
  {Sun}}\ and\ \bibinfo {author} {\bibfnamefont {P.}~\bibnamefont
  {Harrowell}},\ }\href {\doibase 10.1063/5.0094386} {\bibfield  {journal}
  {\bibinfo  {journal} {J. Chem. Phys.}\ }\textbf {\bibinfo {volume} {157}},\
  \bibinfo {pages} {024501} (\bibinfo {year} {2022})}\BibitemShut {NoStop}%
\bibitem [{\citenamefont {Coslovich}\ and\ \citenamefont
  {Ikeda}(2022)}]{coslovichRevisitingSinglesaddleModel2022}%
  \BibitemOpen
  \bibfield  {author} {\bibinfo {author} {\bibfnamefont {D.}~\bibnamefont
  {Coslovich}}\ and\ \bibinfo {author} {\bibfnamefont {A.}~\bibnamefont
  {Ikeda}},\ }\href {\doibase 10.1063/5.0083173} {\bibfield  {journal}
  {\bibinfo  {journal} {J. Chem. Phys.}\ }\textbf {\bibinfo {volume} {156}},\
  \bibinfo {pages} {094503} (\bibinfo {year} {2022})}\BibitemShut {NoStop}%
\bibitem [{\citenamefont {Coslovich}\ \emph {et~al.}(2025)\citenamefont
  {Coslovich}, \citenamefont {Galliano},\ and\ \citenamefont
  {Costigliola}}]{zenodo}%
  \BibitemOpen
  \bibfield  {author} {\bibinfo {author} {\bibfnamefont {D.}~\bibnamefont
  {Coslovich}}, \bibinfo {author} {\bibfnamefont {L.}~\bibnamefont {Galliano}},
  \ and\ \bibinfo {author} {\bibfnamefont {L.}~\bibnamefont {Costigliola}},\
  }\href {\doibase 10.5281/zenodo.14833443} {\enquote {\bibinfo {title}
  {Dataset: Freezing, melting and the onset of glassiness in binary
  mixtures (1.0.1)}}\ } (\bibinfo {year}
  {2025})\BibitemShut {NoStop}%
\end{thebibliography}

%

\end{document}